\documentclass[sigconf]{acmart}%

\usepackage{pgfplots}          %
\pgfplotsset{compat=1.18}      %
\usepgfplotslibrary{statistics,groupplots} %
\definecolor{cb1}{HTML}{0072B2} %
\definecolor{cb2}{HTML}{D55E00}
\definecolor{cb3}{HTML}{009E73}
\definecolor{cb4}{HTML}{CC79A7}

\usepackage{enumitem}

\usepackage[nameinlink, noabbrev, capitalize]{cleveref}%
\usepackage{graphicx} %
\usepackage{orcidlink} %
\usepackage{xcolor}
\usepackage{breakurl}%
\usepackage{natbib}%
\usepackage{tabularx}
\usepackage{booktabs}

\newcommand{\linecont}{\mbox{}\hspace*{1em}}

\PassOptionsToPackage{hyphens}{url}\usepackage{hyperref}%

\copyrightyear{2026}
\acmYear{2026}
\setcopyright{cc}
\setcctype{by}
\acmConference[SIGCSE TS 2026]{Proceedings of the 57th ACM Technical Symposium on Computer Science Education V.1}{February 18--21, 2026}{St. Louis, MO, USA}
\acmBooktitle{Proceedings of the 57th ACM Technical Symposium on Computer Science Education V.1 (SIGCSE TS 2026), February 18--21, 2026, St. Louis, MO, USA}
\acmDOI{10.1145/3770762.3772662}
\acmISBN{979-8-4007-2256-1/26/02}

\newcommand{\erin}[1]{\textcolor{purple} {\textbf{Erin}: #1} }

\title{Owlgorithm: Supporting Self-Regulated Learning in Competitive Programming through LLM-Driven Reflection}
\date{Feb 2026}

\author{Juliana Nieto-Cardenas}
\affiliation{
  \institution{Universidad Nacional de Colombia}
  \city{Bogotá}
  \state{D.C}
  \country{Colombia}
}
\orcid{0009-0009-7881-9810}

\author{Erin Joy Kramer}
\affiliation{
  \institution{Purdue University}
  \city{West Lafayette}
  \state{Indiana}
  \country{USA}
}
\orcid{0009-0002-5918-2382}

\author{Peter Kurto}
\affiliation{
  \institution{Purdue University}
  \city{West Lafayette}
  \state{Indiana}
  \country{USA}
}
\orcid{0009-0007-2430-7700}

\author{Ethan Dickey}
\affiliation{
  \institution{Purdue University}
  \city{West Lafayette}
  \state{Indiana}
  \country{USA}
  }
\email{dickeye@purdue.edu}
\orcid{0009-0007-3706-5253}

\author{Andres Bejarano}
\affiliation{
  \institution{Purdue University}
  \city{West Lafayette}
  \state{Indiana}
  \country{USA}
  }
\email{abejara@purdue.edu}
\orcid{0000-0003-2611-2855}

\begin{abstract}
We present \textit{Owlgorithm}, an educational platform that supports Self-Regulated Learning (SRL) in competitive programming (CP) through AI-generated reflective questions. Leveraging GPT-4o, Owlgorithm produces context-aware, metacognitive prompts tailored to individual student submissions. Integrated into a second- and third-year CP course, the system-provided reflective prompts adapted to student outcomes: guiding deeper conceptual insight for correct solutions and structured debugging for partial or failed ones.

Our exploratory assessment of student ratings and TA feedback revealed both promising benefits and notable limitations. While many %
found the generated questions useful for reflection and debugging, concerns were raised about feedback accuracy and classroom usability. These results suggest advantages of LLM-supported reflection for novice programmers, though refinements are needed to ensure reliability and pedagogical value for advanced learners.

From our experience, several key insights emerged: GenAI can effectively support structured reflection, but careful prompt design, dynamic adaptation, and usability improvements are critical to realizing their potential in education. We offer specific recommendations for educators using similar tools and outline next steps to enhance Owlgorithm's educational impact. The underlying framework may also generalize to other reflective learning contexts.
\end{abstract}

\begin{document}

\ccsdesc[500]{Social and professional topics~Computing education}

\keywords{Competitive Programming, Self-Regulated Learning, Generative AI in CS Education, Reflection, Educational Technology}

\maketitle

\section{Introduction}

Competitive programming (CP) places students in fast, test-driven contests demanding algorithmic solutions under strict time and memory limits. The format naturally supports the planning phase (problems selection, time budgeting) of self-regulated learning (SRL, \cite{zimmerman, pintrich2004framework}) and the performance phase (iterative coding and resubmission to an online judge). Reflection, however, is seldom practiced \cite{loksa2022metacognition}. Once receiving the coveted ``Accepted,'' the scoreboard and time pressure push competitors to the next problem; if the verdict is ``Wrong Answer,'' the rapid trial-and-error culture still discourages systematic postmortems. Online judges provide little scaffolding beyond pass/fail feedback, so revisiting code to trace logic flaws, compare algorithms, or generalize lessons imposes high cognitive and metacognitive costs with no immediate contest payoff \cite{linn_dalbey_1985,alhazbi2014using}.

Within computer science instruction, the reflection stage of SRL often manifests as self-explanation, where students verbalize the logic behind their code, debugging strategies, and algorithm choices. Prior research on self-explanation has shown that prompting learners to justify each step of their problem-solving process can significantly improve understanding and retention \cite{margulieux2016training, vihavainen2015benefits, stone2007integrating}. Moreover, training students to self-explain has been shown to improve cognitive skill acquisition and programming performance \cite{bielaczyc1995training}. Building on this, recent tools have begun to automate reflective support by generating questions about student code \cite{askstudents2022}, yet these are rarely grounded in SRL theory or tailored to novice competitive programmers, who may struggle to diagnose errors, articulate reasoning, or connect solutions to underlying algorithmic principles \cite{metacognitive2024}.

Meanwhile, large language models (LLMs) show near-expert performance across programming tasks like code generation, debugging, and algorithmic reasoning \cite{sarsa2022automatic, Shi2024-zk, Lyu2024-nr}. Their ability to interpret code and generate explanations gives them potential to help the reflection phase of SRL. LLMs can scaffold self-explanation and reduce cognitive effort for meaningful reflection by producing targeted, open-ended questions based on student code.

We present \textit{Owlgorithm}, an LLM tutor supporting reflection in CP. Using GPT-4o, it generates context-aware questions guiding self-explanation for passing or failing code. Grounded in SRL, it maps questions to Bloom's Taxonomy to deepen conceptual understanding. Contributions: (1) a dual-role LLM strategy for question generation and review; (2) a workflow adapting prompts to performance; and (3) an exploratory evaluation of student and TA feedback. \textit{Owlgorithm} bridges the reflection gap in CP, helping learners engage deeply with code via structured, AI-supported reflection.

\section{Background \& Related Work}

\subsection{Self-Regulated Learning (SRL)}
The concept of SRL describes learning as an iterative process consisting of three phases: forethought, performance, and reflection \cite{zimmerman}. Zimmerman defines these phases as goal-setting and planning (forethought), strategy implementation and self-monitoring (performance), and self-evaluation and adaptation (reflection). Pintrich and Zusho later expanded this model to include motivational and metacognitive components across each phase, demonstrating that effective learners continually monitor and adapt their understanding and strategies \cite{pintrich2004framework}. This framework has significant implications for programming education, where students often navigate complex problem-solving cycles that reflect SRL's core phases.

Researchers have explored the use of AI as a tool to support SRL in various scenarios. Salvo \textit{et al.} \cite{Salvo20} proposed a mathematical model that incorporates AI-oriented analysis to assist patients undergoing remote healthcare treatments. Specifically, machine learning analyzes patient data, identifying connections between digital skills, motivation, therapy goals, and behavioral feedback. This approach enables personalized, adaptive SRL plans. The goal is to enhance patient engagement and autonomy by using AI to continuously evaluate and customize Internet of Medical Things (IoMT)-supported self-managed healthcare interventions.

Afzaal \textit{et al.} \cite{Afzaal21} proposed an AI approach based on counterfactual explanations to generate automatic and intelligent recommendations that support students’ SRL in a data-driven manner. This approach builds predictive models using machine learning to forecast students’ performance and utilizes explainable AI techniques to identify specific, actionable changes that students can make to improve their performance. As a result, the AI system predicts outcomes and also guides students on which learning behaviors to adjust in order to achieve their desired academic goals.

Rasdi \textit{et al.}\cite{Rasdi23} explored how insurance agents develop SRL skills to adapt to AI technologies in their work, identifying organizational support mechanisms such as training, self-learning initiatives, knowledge sharing, and adaptation to change. While AI tools such as machine learning, NLP, and RPA are utilized in their tasks, the study focuses on how agents self-regulate their learning to use these tools effectively, rather than AI directly supporting SRL. It concludes that organizational and informal supports are vital to empower workers' SRL in leveraging AI for job performance.

Billman \textit{et al.} \cite{Billman24} analyzed how AI usage, framed by SRL theory, affects the academic performance of Indonesian students, highlighting metacognition, motivation, and behavior. AI tools are found to positively influence SRL factors, with motivation having the strongest impact on academic achievement. However, the study warns about potential student dependency on AI. Overall, AI supports SRL by enhancing metacognitive and motivational processes, yet the paper calls for guided and supervised AI use to avoid undermining students’ independent learning skills.

Kong \textit{et al.} \cite{Kong24} proposed a human-centered learning and teaching framework (HCLTF) that incorporates GenAI tools to enhance K-12 students' SRL through the acquisition of domain knowledge. The framework utilizes AI to support SRL by providing personalized prompts, immediate feedback, and opportunities for self-reflection during the forethought, performance, and reflection phases of learning. In this model, AI acts as an interactive partner while teachers facilitate the process, with the goal of empowering students' attention, engagement, confidence, and satisfaction during SRL.

Yang \textit{et al.} \cite{Yang24} examined how pre-service teachers’ basic need satisfaction (autonomy, competence, relatedness) influences AI literacy in introductory AI courses, highlighting emotional engagement and SRL as key mechanisms. They find SRL significantly predicts AI literacy and moderates the relationship between emotional engagement and AI literacy, enhancing learning outcomes when students effectively regulate learning. Thus, SRL emerges as a crucial factor amplifying the positive effects of emotional engagement in acquiring AI literacy within AI-supported educational contexts.

Recently, Ren \textit{et al.} \cite{Ren25} systematically reviewed 27 empirical studies from 2004 to 2024 on how AI supports SRL in education. It finds that AI is primarily used through intelligent tutoring systems, machine learning for predicting learning behavior, educational robots, and AI-enabled learning environments to enhance metacognition, motivation, and behavioral regulation. Overall, AI tools are shown to support SRL by providing personalized feedback, adaptive scaffolding, and predictive analytics, with most studies reporting positive impacts on students’ SRL skills and academic performance.

\subsection{Tools Supporting Reflection}
In CS education, SRL reflection typically appears as self-explanation -- students verbalize code logic, debugging steps, or algorithmic choices. Early work by Linn and Dalbey found that, without scaffolding, students skip this phase and treat programming as a syntactic task \cite{linn_dalbey_1985}. Later studies confirmed the value of self-explanation: relational code explanations correlate with proficiency \cite{explain2012code} and stronger SRL habits predict higher quality code \cite{castellanos2017understanding}.

\subsection{LLMs in Programming Education}
LLMs now automate explanations, feedback, and code generation for programming education. Sarsa \textit{et al.} showed that OpenAI Codex can create programming exercises and explanations rivaling instructor work, indicating potential for scalable instructional support \cite{sarsa2022automatic}. Lyu \textit{et al.} found LLMs achieve near-expert results on complex algorithmic tasks, and Shi \textit{et al.} reported strong performance on Olympiad problems with careful prompting \cite{Lyu2024-nr, Shi2024-zk}. These studies suggest that LLMs can comprehend advanced programming concepts and can assist learners across difficulty and ability levels.

Alongside advancements in LLMs, several educational tools have aimed to support SRL through automation. Alhazbi’s e-journaling system prompted students to reflect on their planning and coding strategies by writing a weekly e-journal entry, leading to measurable gains in metacognitive awareness and academic performance \cite{alhazbi2014using}. Bakry \textit{et al.} developed a system that automatically generates code-specific reflection questions, enabling deeper student engagement without adding to instructor burden \cite{askstudents2022}. However, as Thompson \textit{et al.} point out, while LLMs can generate reflective prompts on demand, few tools have integrated them into structured, adaptive workflows explicitly grounded in SRL theory and aligned with student performance data \cite{metacognitive2024}.

Broader AI-based frameworks such as GAIDE (Generative AI for Instructional Design and Evaluation) have concentrated on automating static content development--such as quizzes, lecture slides, and grading rubrics--but omit SRL-driven interaction \cite{dickey2024GAIDE, kaplan2023generative}. While these systems effectively lessen instructor workload, they do not actively facilitate student reflection, metacognitive reasoning, or adaptive self-regulation during the learning process.

Overall, although LLMs and SRL-aligned prompting systems have made significant progress, no existing tool completes the SRL cycle in the context of competitive programming. Most current systems focus on static content generation or provide generic feedback, with few leveraging LLMs to support structured, performance-sensitive reflection. \textit{Owlgorithm} addresses this gap by combining LLM-driven question generation with SRL-informed design principles, helping students in competitive programming complete the cycle of planning, performance, and meaningful reflection.

\section{Owlgorithm}
\textit{Owlgorithm} is a web-based platform designed to support self--explanation and reflective practice in competitive programming by generating tailored questions using GPT-4o. The system adjusts its prompts based on student outcomes, distinguishing between two modes: submissions that pass all test cases and those that fail some or all test cases. In the former, the questions emphasize algorithmic generalization and complexity; in the latter, they facilitate debugging and self-diagnosis.

\subsection{SRL and Programming Concepts}
SRL is a metacognitive cycle in which learners actively control their progress by setting goals, employing strategies, and reflecting on outcomes \cite{zimmerman}. In Zimmerman’s model, the forethought phase involves task analysis and goal setting: for example, a student may decide to master segment tree problems by solving two per session and allocating 60 minutes for each. During the performance phase, learners monitor and manage their behavior using techniques such as incremental coding, debug prints to verify partial results, and time-boxing to avoid Time Limit Exceeded verdicts. Finally, the self-reflection phase requires learners to interpret feedback by examining failed test cases, attributing errors (e.g., off-by-one logic, incorrect data structures), and adjusting future study strategies, such as reviewing prefix-sum techniques.

SRL encourages iterative improvement by prompting learners to revise strategies based on outcomes, seek external resources when necessary, and reflect on patterns of success and failure \cite{loksa2022metacognition}. However, as prior work showed, many students entirely skip the reflection phase, finding it cognitively demanding and unstructured, especially in fast-paced environments such as competitive programming.

Clear learning objectives are essential to guide this reflective process. The Revised Bloom’s Taxonomy provides a structured framework for classifying educational goals across cognitive levels, from remembering and understanding to evaluating and creating. Unlike the original taxonomy, this taxonomy includes metacognitive knowledge as a distinct dimension, emphasizing the importance of learners’ awareness of their own thinking processes \cite{conklin2005taxonomyreview}. As such, the Revised Bloom's Taxonomy is widely used in the literature to support the design of activities that foster self-reflection by explicitly targeting metacognitive development \cite{prather2020metacognition}.

\textit{Owlgorithm} operates under two usage scenarios, selected based on the outcome of a code submission: ``All Cases Passed'' and ``Some or None Cases Passed''. In the All Cases scenario (where a student’s code passes every test case), the system prompts deeper reflection by generating questions focused on generalization, algorithmic justification, and complexity analysis. For example, students may be asked to explain why their solution handles edge cases, justify their choice of algorithm or data structure, or analyze whether the time and space complexity is appropriate for the problem constraints.

In contrast, the Some or None scenario is triggered when a submission fails at least one test due to a Wrong Answer, Time Limit Exceeded, Runtime Error, or Compile Error. In this case, \textit{Owlgorithm} guides students through reflective debugging by prompting them to identify the failure mode, recognize common pitfalls (such as off-by-one errors or input parsing bugs), and formulate a revision plan. This approach transforms debugging into a metacognitive process, encouraging root cause analysis and strategic problem-solving rather than trial-and-error iteration.

\subsection{High Level Architecture}
\subsubsection{Two LLM Roles} \textit{Owlgorithm} separates LLM behavior into two distinct roles, Generator and Reviewer (two distinct OpenAI API client instances), to maintain clarity and control in the reflection question pipeline. The Generator role creates candidate questions and exemplar answers based on the student’s code and problem context, using prompt templates aligned with specific learning objectives. These outputs focus on core concepts such as correctness, algorithmic complexity, and design rationale. The Reviewer role then evaluates these questions for clarity, cognitive depth, and alignment with Bloom’s Taxonomy. It filters out low-relevance or redundant items and adjusts questions to match the intended difficulty level. This separation helps prevent context bleeding, a failure mode in LLM systems where overlapping goals or instructions cause incoherent or unfocused outputs, as discussed in the GAIDE architecture \cite{dickey2024GAIDE}. This two-step process ensures that reflective prompts remain pedagogically targeted, cognitively appropriate, and aligned with SRL principles by decoupling generation from evaluation.

\subsubsection{Prompt Flow} Each usage scenario in \textit{Owlgorithm} follows a structured prompt pipeline designed to scaffold both question generation and evaluation. In the All Cases scenario (where the student’s code passes all test cases), the Generator produces open-ended prompts that target higher-order reflection, focusing on abstraction, justification, and complexity analysis. Example questions include: \textit{``Why does your algorithm handle edge case X correctly?''} and \textit{``Explain the time complexity of your solution.''}.%

The Reviewer then selects a focused subset of these questions, adds exemplar responses, and formulates rubrics aligned with Bloom’s Taxonomy to ensure pedagogical consistency. These rubrics are intended to support self-assessment and instructional feedback, helping learners evaluate correctness and depth of understanding.

In contrast, the Some or None Cases scenario, where the student’s code fails one or more test cases, shifts the focus to error diagnosis and revision planning. In this scenario, the prompt flow adapts to focus on diagnosis and revision. Questions in this mode are more targeted,  that ask students to identify likely error categories, reassess edge case handling, or question initial assumptions.

Each prompt targets common failure patterns such as logic flaws, indexing mistakes, or type mismatches, and encourages students to develop a revision plan rather than rely on ad hoc debugging. These diagnostic prompts align with the reflection phase of SRL by fostering self-monitoring, error attribution, and strategic adjustment. Prompt flows across both success and failure scenarios are designed to maintain cognitive variety and progression, scaffolding students from surface-level issues to deeper conceptual understanding.

\subsection{System Workflow}
\textit{Owlgorithm} guides students through a concise, five-step workflow that scaffolds self-reflection with minimal overhead. A simple landing page launches the process, followed by file upload (problem statement and code) and session setup, where students indicate whether their submission passes or fails test cases and, if failing, select a verdict (e.g., Wrong Answer, Runtime Error, Time Limit Exceeded). For unique code/problem pairs, GPT-4o generates Bloom-aligned reflection questions, scores student responses (0-3), and provides concise formative feedback; failing code also receives sub-50-word hints toward likely faults. A final summary compiles all questions, responses, and feedback--helping students identify patterns and consolidate insights. The entire process takes under five minutes, lowering the barrier to deliberate, theory-informed reflection during or after contests.

\subsection{Detailed Prompt Flow}
We found that prompt chaining and having multiple GPT agents working together was very effective at reducing errors to near-zero and keeping content higher quality. Errors that did occur (anything not in the specified regex format) were immediately regenerated -- with high second-time success rate.
\subsubsection{All cases scenario}
\begin{figure*}
  \centering
  \includegraphics[width=\linewidth]{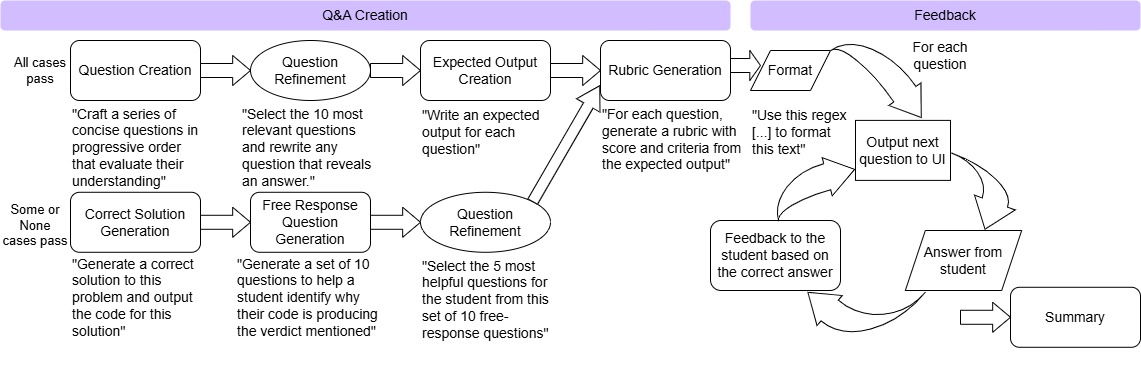}
  \caption{Workflow of the prompts used in both ``All'' and ``Some or None'' Cases Passed. Rounded corners show AI generation, circles show iterative AI refinement, rhombuses show formatting text for processing, sharp-cornered boxes show UI display.}
  \Description{A diagram showing the workflow of prompts for the All Cases Pass scenario.}
  \label{fig:owl_workflow}
\end{figure*}

\Cref{fig:owl_workflow} shows the end-to-end prompt chaining used in to generate self-reflection questions and give feedback when the student’s code passes all test cases.

\paragraph{Question Creation}
\begin{enumerate}[leftmargin=*]
    \item \textbf{Generator -- draft.} OpenAI API client $\# 1$ creates $\approx 20$ open-ended questions from the problem and code.
    \item \textbf{Reviewer -- refine.} OpenAI API client $\# 2$ filters and rephrases them to the 10 most pedagogically useful.
    \item \textbf{Generator -- answers.} $\# 1$ writes expected answers for each refined question.
    \item \textbf{Reviewer -- rubric.} $\# 2$ builds scoring criteria mapped to Bloom’s Taxonomy levels.
    \item \textbf{Formatter -- package.} OpenAI API client $\# 3$ Outputs questions, answers, and rubrics in JSON/Markdown for the UI. We have found LLMs capable of generating highly specific formats well enough to rely on them without being able to verify.
\end{enumerate}

\paragraph{Feedback generation}
A feedback instance (OpenAI API client instance) invokes the feedback prompt passing in the student’s answer, the original question, problem statement, code, and the rubric, to compare the response against each rubric criterion. It assigns a numeric score (0-3) based on relevance and correctness, then generates a concise hint (20 words maximum) without revealing the full solution. The output is formatted to match a regex.

\paragraph{Summary generation}
A summary maker instance (OpenAI API client instance) applies the summary prompt to a list of question dictionaries (each containing a ‘Question Statement’, ‘Score’, and ‘Feedback message’) to generate concise question titles ($\leq$ 5 words) and an HTML summary table with columns for Question Title, Score, and Feedback messages.

\subsubsection{None or Some Cases Scenario}
\Cref{fig:owl_workflow} shows the end-to-end prompt chaining used to generate diagnostic questions and provide feedback when the student's code does not pass all test cases.

\paragraph{Question Creation}
\begin{enumerate}[leftmargin=*]
    \item \textbf{Generator -- reference.} OpenAI API client $\# 1$ produces a correct Python solution as baseline.
    \item \textbf{Generator -- questions.} $\# 1$ crafts 10 free-response prompts probing logic gaps.
    \item \textbf{Reviewer -- select.} OpenAI API client $\# 2$ chooses the 5 most helpful, verdict-aligned questions.
    \item \textbf{Reviewer -- rubric.} $\# 2$ builds Bloom-aligned scoring criteria.
    \item \textbf{Formatter -- package.} OpenAI API client $\# 3$ Outputs questions + rubric as JSON for the UI.
\end{enumerate}

\paragraph{Feedback Generation}
A feedback invokes the feedback prompt, passing in the student's answer, the original question, problem statement, code, and the rubric. It compares the response against each rubric criterion, assigns a numeric score (0-3) based on relevance and correctness, and generates a concise hint without revealing the full solution. The output is formatted to match a complicated regex, which the LLM matches remarkably well.

\subsubsection{Prompt Engineering Techniques}
To ensure consistency, relevance, and pedagogical alignment, the following prompt-engineering strategies are applied at each stage:

\begin{itemize}[leftmargin=*]
  \item \textbf{Role Specification.}  
    Each prompt fixes a persona (e.g., ``Competitive Programming Professor'') via the \textit{persona prompt pattern} \cite{white2023promptpattern}, to control tone, style, and expertise, producing more focused, relevant, and immersive responses.

  \item \textbf{Structured Delimiters.}  
    Tags like \verb|<problem>| and \verb|<code>| mark sections that carrying over context from previous prompts, enabling seamless prompt chaining \cite{wu2021aichains}. 

  \item \textbf{Incremental Refinement.}
    Generator + reviewer chaining iteratively sharpens question clarity and relevance.
  \item \textbf{Regex Enforcement.}  
    Regex in the prompt enforce output formats (lists, headings, JSON fields) for easy parsing.
  \item \textbf{Bloom’s Taxonomy Verbs.}  
    Prompts for expected outputs and rubrics incorporate cognitive-level verbs (e.g., Remember, Understand, Apply) to target diverse learning objectives.
  \item \textbf{Temperature Control.}
    Low temperature keeps outputs deterministic, yielding reproducible questions and rubrics.
  \item \textbf{Explicit Scoring Guidelines.}  
    Rubrics use 0-3 scales with descriptive anchors for consistent feedback and hints.
\end{itemize}

\subsection{Design Rationale}
\textit{Owlgorithm} is purpose-built to operationalize the \emph{self-reflection} phase of Zimmerman’s Self-Regulated Learning (SRL) cycle--the stage in which learners interpret outcomes, attribute causes, and revise subsequent strategies~\cite{zimmerman}. Designed for the Competitive Programming context, it aims to help students close the SRL loop by transforming post-submission analysis into a structured, reflective learning opportunity. After each code submission, a two-step LLM pipeline--\textit{Generator} followed by \textit{Reviewer}--produces Bloom-stratified prompts that guide learners from surface-level recall toward progressively higher-order cognitive processes: \emph{Remember}, \emph{Understand}, \emph{Apply}, \emph{Analyze}, and \emph{Integrate}~\cite{conklin2005taxonomyreview}. These prompts are tailored to core Competitive Programming skills like algorithmic reasoning, debugging, and complexity optimization, fostering deeper metacognitive appraisal of problem-solving strategies. By grounding each prompt in concrete artifacts--line numbers, variable states, and the online judge’s verdict--the system situates abstract reflection verbs within the learner’s own code, leveraging evidence that higher-level Bloom prompts enhance programming comprehension~\cite{thompson2008blooms}. Because reflection is fully automated, instructors incur virtually no additional workload, while learners receive timely, cognitively rich feedback precisely within the brief post-submission window when reflective motivation peaks.

\section{Usage and Perception Analysis}
Each question generated by \textit{Owlgorithm} permits the user to rate how helpful the question and feedback to the students' response is. From this 5-star rating system, it became clear that significant improvements to question and feedback quality were still needed.

In Spring 2025, we conducted a perception survey of \textit{Owlgorithm} among six Undergraduate Teaching Assistants (UTAs) and one instructor, representing the majority of instructional staff for our Competitive Programming courses CP1, CP2, and CP3.
All of the UTAs have assisted with CP1, four with CP2, and three with CP3. 

The survey consisted of a 25-item questionnaire that took approximately 5 to 10 minutes to complete. It explored seven key areas: the quality of generated questions, accuracy and utility of feedback, expectations for student adoption, overall helpfulness, debugging support, anticipated learning gains, and space for open-ended comments. A summary is provided in \Cref{table:survey_data}.

\begin{table*}[ht!]
    \centering
  \setlength{\tabcolsep}{4pt}      %
    \begin{tabularx}{\textwidth}{@{}p{0.14\textwidth} @{}p{0.22\textwidth} X X}
        \toprule
        \textbf{Dimension} & \textbf{Key Metrics} & \textbf{Summary of Ratings} & \textbf{Notable Observations} \\
        \midrule
        \textbf{Question \newline Quality} &
        Correctness: 21--100\%,      \newline \linecont M = 62.1\%, SD = 25.8\% \newline
        Helpfulness (Understanding): \newline \linecont M = 3.0, SD = 0.8 \newline
        Helpfulness (Debugging):     \newline \linecont M = 3.4, SD = 1.0 &
        Neutral to slightly positive ratings \newline
        Moderate perceived debugging value &
        Wide variation in correctness \newline
        Higher perceived value for debugging vs. understanding \\
        \midrule
        \textbf{Feedback \newline Quality} &
        Correctness: 19--100\%,      \newline \linecont M = 49.9\%, SD = 30.4\% \newline
        Helpfulness (Understanding): \newline \linecont M = 2.7, SD = 1.0 \newline
        Helpfulness (Debugging):     \newline \linecont M = 2.9, SD = 0.7 &
        Generally neutral to negative \newline
        Lower than question quality across \newline all metrics &
        High variability; concerns about misleading or incorrect feedback \\
        \midrule
        \textbf{Predicted \newline Student Usage} &
        Debugging: 87.5\% \newline
        CP1 Interviews: 62.5\% &
        Most common use: debugging \newline
        Limited expected frequency across courses &
        CP3: 100\% predicted never use \newline
        CP2: 75\% predicted never use \\
        \midrule
        \textbf{Perceived \newline Helpfulness} &
        CP1: M = 3.1, SD = 0.8 \newline
        CP2: M = 1.9, SD = 0.6 \newline
        CP3: M = 1.7, SD = 0.5 &
        Slightly positive (CP1) \newline
        Negative (CP2, CP3) &
        Perceived value drops sharply with course level \\
        \midrule
        \textbf{Debugging \newline Usefulness} &
        CP1: M = 3.3, SD = 0.7 \newline
        CP2: M = 2.3, SD = 0.7 \newline
        CP3: M = 2.0, SD = 0.6 &
        Moderate for CP1 \newline
        Low for CP2, CP3 &
        Best suited for lower-level courses \\
        \midrule
        \textbf{Anticipated \newline Learning \newline Impact} &
        CP1: M = 3.0, SD = 0.0 \newline
        CP2: M = 2.1, SD = 0.6 \newline
        CP3: M = 1.7, SD = 0.5 &
        Neutral (CP1) \newline
        Negative learning impact (CP2, CP3) &
        Concern about incorrect or unhelpful learning at higher levels \\
        \midrule
        \textbf{Open Feedback \newline \& Limitations} &
        -- &
        During first round, TAs noted hallucinations, poor feedback quality, UI issues; subsequent rounds had notably higher ratings &
        Tool seen as useful mainly for CP1 \newline
        Criticism of non-interactive format \newline
        Preference for tools like ChatGPT \\
        \bottomrule
    \end{tabularx}
    \caption{Summary of TA Evaluations of \textit{Owlgorithm} by Category}
    \label{table:survey_data}
\end{table*}

TAs in this study expressed mixed perceptions regarding \textit{Owlgorithm}’s effectiveness for student learning, with notable differences across course levels. In CP1, TAs demonstrated modest optimism; half anticipated that students would engage with the tool weekly, and all agreed that it would support at least minimal learning gains, particularly in debugging and interview preparation. TAs especially valued \textit{Owlgorithm}’s potential to complement or partially replace office hour support in straightforward contexts by providing troubleshooting opportunities for students who are failing test cases.

In contrast, this optimism was not observed in higher-level courses. In CP2, five out of eight TAs predicted no student adoption, and in CP3, seven out of eight anticipated no usage. TAs described \textit{Owlgorithm} as ``office-hours on demand,'' highlighting its ability to extend guided support beyond scheduled sessions, direct novices toward root causes rather than complete solutions, reduce last-minute stress, and offer post-solution conceptual deep dives for advanced learners preparing for interviews or oral examinations.

However, TAs also identified limitations that constrain the tool’s value, particularly in upper-division courses. These include intermittent hallucinations and scoring errors, slow response times, and the absence of a conversational, adaptive interface. Collectively, these issues underscore the need for improvements in accuracy, performance, and interactivity before the tool can be widely adopted.

In exploratory testing, surveyed students and in-app ratings suggest that students appreciated many of the ``high-quality'' prompts \textit{Owlgorithm} produced, yet just under half still reported low satisfaction. Their chief frustration was generic hypotheticals--e.g., ``How would your code act in <insert certain scenario>?''--that repeated scenarios already covered, along with broad edge-case questions instead of bug-specific guidance. Consequently, only about half the questions felt useful. The issue traces to the LLM's limited context window: as inputs grow, it loses track of crucial code details. Splitting work into smaller queries helped somewhat, but further refinement is needed for scaled, cost-effective framework use.

\section{Discussion \& Future Work}

Our pilot (mostly with TAs and limited CP1 initial deployment) shows that even a lightweight version of \textit{Owlgorithm} can spark productive self-reflection: roughly half of its prompts were rated ``high-quality,'' and several students reported clearer reasoning after one session. Still, limited feedback accuracy ($\approx40\%$), interface delays, and a fixed question flow sometimes frustrated users, especially when the AI repeated scenarios they had already handled.

Future work should improve feedback precision using larger-context or retrieval-augmented LLMs (or other cost-effective ``in-place'' adjustments), refine the UI based on pilot feedback, and replace the static prompt list with a conversational engine that adapts to each answer. These refinements will position \textit{Owlgorithm} to deliver TA-like guidance at scale: a practical, low-friction coach for novice competitive programmers and other iterative CS courses.

In classrooms, \textit{Owlgorithm} can serve as an office-hour–adjacent or post-contest reflection tool that integrates naturally into existing contest workflows. Its backend can be readily swapped for alternative or institutionally approved LLMs, ensuring continued use even where GPT-4o access is restricted. Since online judges already support file-based code submissions and downloadable problem statements, adoption requires no extra steps for instructors or students. A practical adoption roadmap includes: (1) guided reflection sessions in early labs, (2) optional self-reflection check-ins tied to contest submissions, and (3) LMS integration for automated prompt delivery and progress tracking. Together, these steps support scalable classroom use while preserving its core goal—lowering the barrier to deliberate, theory-informed reflection during or after contests.
We conclude with an outlook shared by the seasoned CP1 instructor, who provided feedback outside of exploratory testing:

\textit{``Owlgorithm represents a new era in [CP] education. It provides direct support for student learning by guiding them using [similar questions] to the ones I use... boiling down issues to core misunderstandings. ...
The primary benefit of this tool is to help the struggling students, the frustrated ones who are sitting at their computer the night of the deadline and just can't do it, can't think anymore, ... and are all alone. The ones who just need to talk to a TA, a friend... or the instructor... about their problems. Someone to... point them in the right direction, because they're capable of doing the problem themselves but [can't figure out the] next step. I look forward to sharing this with my students next semester!''}

\section*{Acknowledgments}
This work was funded by Purdue's Innovation Hub (IH-AI-23002) and the Department of Computer Science through the GoBoiler program. The authors also acknowledge the Teaching Assistants from the CS211/311/411 course who participated in the pilot program for \textit{Owlgorithm} during the Spring of 2025.

\bibliographystyle{ACM-Reference-Format}
\bibliography{refs}

@article{kaplan2023generative,
  title={Generative AI and teachers’ perspectives on its implementation in education},
  author={Kaplan-Rakowski, Regina and Grotewold, Kimberly and Hartwick, Peggy and Papin, Kevin},
  journal={Journal of Interactive Learning Research},
  volume={34},
  number={2},
  pages={313--338},
  year={2023},
  publisher={Association for the Advancement of Computing in Education (AACE)},
url     = {https://www.learntechlib.org/primary/p/222363/}
}

@inproceedings{dickey2024GAIDE,
    author={Dickey, Ethan and Bejarano, Andres},
    booktitle={2024 IEEE Frontiers in Education Conference (FIE)}, 
    title={GAIDE: A Framework for Using Generative AI to Assist in Course Content Development}, 
    year={2024},
    publisher    = {IEEE},
    address      = {New York, NY, USA},
    pages={1-9},
    doi={10.1109/FIE61694.2024.10893132}
}

@article{zimmerman,
    author    = {Zimmerman, Barry J.},
    title     = {Becoming a Self-Regulated Learner: An Overview},
    journal   = {Theory Into Practice},
    volume = {41},
    number = {2},
    pages = {64--70},
    year = {2002},
    publisher = {Routledge},
    doi = {10.1207/s15430421tip4102\_2},
    URL = {https://doi.org/10.1207/s15430421tip4102_2},
}

@article{pintrich2004framework,
    author    = {Paul R. Pintrich},
    title     = {A Conceptual Framework for Assessing Motivation and Self-Regulated Learning in College Students},
    journal   = {Educational Psychology Review},
    year      = {2004},
    volume    = {16},
    number    = {4},
    pages     = {385--407},
    doi       = {10.1007/s10648-004-0006-x},
    url       = {https://doi.org/10.1007/s10648-004-0006-x},
    issn      = {1573-336X},
    publisher = {Springer},
    address   = {New York, NY, USA}
}

@inproceedings{sarsa2022automatic,
    author = {Sarsa, Sami and Denny, Paul and Hellas, Arto and Leinonen, Juho},
    title = {Automatic Generation of Programming Exercises and Code Explanations Using Large Language Models},
    year = {2022},
    isbn = {9781450391948},
    publisher = {Association for Computing Machinery},
    address = {New York, NY, USA},
    url = {https://doi.org/10.1145/3501385.3543957},
    doi = {10.1145/3501385.3543957},
    booktitle = {Proceedings of the 2022 ACM Conference on International Computing Education Research - Volume 1},
    pages = {27–43},
    numpages = {17},
    location = {Lugano and Virtual Event, Switzerland},
    series = {ICER '22}
}

@inproceedings{explain2012code,
    author = {Murphy, Laurie and Fitzgerald, Sue and Lister, Raymond and McCauley, Ren\'{e}e},
    title = {Ability to 'explain in plain english' linked to proficiency in computer-based programming},
    year = {2012},
    isbn = {9781450316040},
    publisher = {Association for Computing Machinery},
    address = {New York, NY, USA},
    url = {https://doi.org/10.1145/2361276.2361299},
    doi = {10.1145/2361276.2361299},
    booktitle = {Proceedings of the Ninth Annual International Conference on International Computing Education Research},
    pages = {111–118},
    numpages = {8},
    location = {Auckland, New Zealand},
    series = {ICER '12}
}

@article{bielaczyc1995training,
  author = {Katerine Bielaczyc and Peter L. Pirolli and Ann L. Brown},
 journal = {Cognition and Instruction},
 number = {2},
 pages = {221--252},
 publisher = {Taylor & Francis, Ltd.},
 title = {Training in Self-Explanation and Self-Regulation Strategies: Investigating the Effects of Knowledge Acquisition Activities on Problem Solving},
 
 volume = {13},
 year = {1995},
  doi          = {10.1207/s1532690xci1302\_3}
}

@article{loksa2022metacognition,
    author       = {Loksa, Dastyni and Margulieux, Lauren and Becker, Brett A. and Craig, Michelle and Denny, Paul and Pettit, Raymond and Prather, James},
    title        = {Metacognition and Self-Regulation in Programming Education: Theories and Exemplars of Use},
    journal      = {ACM Transactions on Computing Education},
    volume       = {22},
    number       = {4},
    pages        = {39:1--39:31},
    year         = {2022},
    doi          = {10.1145/3487050},
    url = {https://doi.org/10.1145/3487050},
}

@inproceedings{vihavainen2015benefits,
    author = {Vihavainen, Arto and Miller, Craig S. and Settle, Amber},
    title = {Benefits of Self-explanation in Introductory Programming},
    year = {2015},
    isbn = {9781450329668},
    publisher = {Association for Computing Machinery},
    address = {New York, NY, USA},
    url = {https://doi.org/10.1145/2676723.2677260},
    doi = {10.1145/2676723.2677260},
    booktitle = {Proceedings of the 46th ACM Technical Symposium on Computer Science Education},
    pages = {284–289},
    numpages = {6},
    location = {Kansas City, Missouri, USA},
    series = {SIGCSE '15}
}

@inproceedings{prather2020metacognition,
  author       = {James Prather and Brett A. Becker and Michelle Craig and Paul Denny and Dastyni Loksa and Lauren Margulieux},
  title        = {What Do We Think We Think We Are Doing?: Metacognition and Self-Regulation in Programming},
  booktitle    = {Proceedings of the 2020 ACM Conference on International Computing Education Research (ICER '20)},
  year         = {2020},
  pages        = {2--13},
  publisher    = {Association for Computing Machinery},
  address      = {Virtual Event, New Zealand},
  doi          = {10.1145/3372782.3406263},
  isbn         = {978-1-4503-7092-9}
}

@article{stone2007integrating,
  author       = {Stone, Jeffrey A. and Madigan, Elinor M.},
  title        = {Integrating Reflective Writing in CS/IS},
  journal      = {SIGCSE Bulletin},
  volume       = {39},
  number       = {2},
  pages        = {42--45},
  year         = {2007},
  doi          = {10.1145/1272848.1272881}
}

@inproceedings{thompson2008blooms,
  author    = {Errol Thompson and Andrew Luxton-Reilly and Jacqueline L. Whalley
               and Minjie Hu and Phil Robbins},
  title     = {Bloom's Taxonomy for CS Assessment},
  booktitle = {Proceedings of the Tenth Australasian Computing Education
               Conference (ACE 2008)},
  series    = {CRPIT},
  volume    = {78},
  pages     = {155--162},
  year      = {2008},
  publisher = {Australian Computer Society},
  address   = {Wollongong, Australia},
  
}

@article{conklin2005taxonomyreview,
  author       = {Conklin, Jack},
  title        = {Review of *A Taxonomy for Learning, Teaching, and Assessing: A Revision of Bloom’s Taxonomy of Educational Objectives—Complete Edition* by Lorin W. Anderson et al.},
  journal      = {Educational Horizons},
  volume       = {83},
  number       = {3},
  pages        = {154--159},
  year         = {2005},
  note         = {Spring 2005 issue, “The Quest for Strengths: The Dawn of a Talent‑Based Approach to K‑12 Education”}
}

@inproceedings{castellanos2017understanding,
    author={Castellanos, Hugo and Restrepo-Calle, Felipe and González, Fabio A. and Echeverry, Jhon Jairo Ramírez},
    booktitle={IEEE Frontiers in Education Conference (FIE)}, 
    title={Understanding the relationships between self-regulated learning and students source code in a computer programming course}, 
    year={2017},
    publisher    = {IEEE},
    address      = {New York, NY, USA},
    pages={1-9},
    doi={10.1109/FIE.2017.8190467}
}

@inproceedings{alhazbi2014using,
    author={Alhazbi, Saleh},
    booktitle={IEEE Global Engineering Education Conference (EDUCON)}, 
    title={Using e-journaling to improve self-regulated learning in introductory computer programming course}, 
    year={2014},
    publisher    = {IEEE},
    address      = {New York, NY, USA},
    pages={352-356},
    doi={10.1109/EDUCON.2014.6826116}
}

@inproceedings{metacognitive2024,
    author = {Tankelevitch, Lev and Kewenig, Viktor and Simkute, Auste and Scott, Ava Elizabeth and Sarkar, Advait and Sellen, Abigail and Rintel, Sean},
    title = {The Metacognitive Demands and Opportunities of Generative AI},
    year = {2024},
    isbn = {9798400703300},
    publisher = {Association for Computing Machinery},
    address = {New York, NY, USA},
    url = {https://doi.org/10.1145/3613904.3642902},
    doi = {10.1145/3613904.3642902},
    booktitle = {Proceedings of the 2024 CHI Conference on Human Factors in Computing Systems},
    articleno = {680},
    numpages = {24},
    location = {Honolulu, HI, USA},
    series = {CHI '24}
}

@inproceedings{margulieux2016training,
  author    = {Lauren E. Margulieux and Briana B. Morrison and Mark Guzdial and Richard Catrambone},
  title     = {Training Learners to Self-Explain: Designing Instructions and Examples to Improve Problem Solving},
  booktitle = {Transforming Learning, Empowering Learners: Proceedings of the 12th International Conference of the Learning Sciences (ICLS 2016), Volume 1},
  pages     = {98--105},
  address   = {Singapore},
  year      = {2016},
  publisher = {International Society of the Learning Sciences (ISLS)},
  url       = {https://repository.isls.org/handle/1/104}
}

@inproceedings{askstudents2022,
    author={Lehtinen, Teemu and Santos, André L. and Sorva, Juha},
    booktitle={2021 IEEE/ACM 29th International Conference on Program Comprehension (ICPC)}, 
    title={Let’s Ask Students About Their Programs, Automatically}, 
    year={2021},
    publisher    = {IEEE},
    address      = {New York, NY, USA},
    pages={467-475},
    doi={10.1109/ICPC52881.2021.00054}
}

@misc{Lyu2024-nr,
    title={Large Language Models as Code Executors: An Exploratory Study}, 
    author={Chenyang Lyu and Lecheng Yan and Rui Xing and Wenxi Li and Younes Samih and Tianbo Ji and Longyue Wang},
    year={2024},
    eprint={2410.06667},
    archivePrefix={arXiv},
    primaryClass={cs.CL},
    url={https://arxiv.org/abs/2410.06667}, 
}

@misc{Shi2024-zk,
    title={Can Language Models Solve Olympiad Programming?}, 
    author={Quan Shi and Michael Tang and Karthik Narasimhan and Shunyu Yao},
    year={2024},
    eprint={2404.10952},
    archivePrefix={arXiv},
    primaryClass={cs.CL},
    url={https://arxiv.org/abs/2404.10952}, 
}

@article{linn_dalbey_1985,
  author  = {Linn, Marcia C. and Dalbey, John},
  title   = {Cognitive Consequences of Programming Instruction: Instruction, Access, and Ability},
  journal = {Educational Psychologist},
  year    = {1985},
  volume  = {20},
  number  = {4},
  pages   = {191--206},
  month   = {Sep},
  
}

@misc{white2023promptpattern,
    author={Jules White and Quchen Fu and Sam Hays and Michael Sandborn and Carlos Olea and Henry Gilbert and Ashraf Elnashar and Jesse Spencer-Smith and Douglas C. Schmidt},
    title={A Prompt Pattern Catalog to Enhance Prompt Engineering with ChatGPT}, 
    year={2023},
    eprint={2302.11382},
    archivePrefix={arXiv},
    primaryClass={cs.SE},
    url={https://arxiv.org/abs/2302.11382}, 
}

@misc{wu2021aichains,
    title={AI Chains: Transparent and Controllable Human-AI Interaction by Chaining Large Language Model Prompts}, 
    author={Tongshuang Wu and Michael Terry and Carrie J. Cai},
    year={2022},
    eprint={2110.01691},
    archivePrefix={arXiv},
    primaryClass={cs.HC},
    url={https://arxiv.org/abs/2110.01691}, 
}

@inproceedings{Rasdi23,
    author={Rasdi, Roziah Mohd and Umar Baki, Nordahlia and Rasdi, Roslina Mohd},
    booktitle={2023 World Engineering Education Forum - Global Engineering Deans Council (WEEF-GEDC)}, 
    title={Exploring Self-regulated Learning Support for AI Related Work in an Insurance Firm}, 
    year={2023},
    publisher    = {IEEE},
    address      = {New York, NY, USA},
    pages={1-7},
    doi={10.1109/WEEF-GEDC59520.2023.10343803}
}

@inproceedings{Billman24,
    author={Billman, Abdullah and Surjandy},
    booktitle={2024 International Conference on Computer Engineering, Network, and Intelligent Multimedia (CENIM)}, 
    title={Impact Analysis of the Use of Artificial Intelligence with Self-Regulated Learning Theory on Student Academic Performance in Indonesia}, 
    year={2024},
    publisher    = {IEEE},
    address      = {New York, NY, USA},
    pages={1-5},
    doi={10.1109/CENIM64038.2024.10882800}
}

@ARTICLE{Kong24,
  author={Kong, Siu-Cheung and Yang, Yin},
  journal={IEEE Transactions on Learning Technologies}, 
  title={A Human-Centered Learning and Teaching Framework Using Generative Artificial Intelligence for Self-Regulated Learning Development Through Domain Knowledge Learning in K–12 Settings}, 
  year={2024},
  volume={17},
  number={},
  pages={1562-1573},
  doi={10.1109/TLT.2024.3392830}}

@ARTICLE{Ren25,
  author={Ren, Liling and Lee, Kerry and May, Lawrence},
  journal={IEEE Access}, 
  title={A Systematic Review Exploring AI’s Role in Self-Regulated Learning Within Education Contexts}, 
  year={2025},
  volume={13},
  number={},
  pages={109771-109782},
  doi={10.1109/ACCESS.2025.3582600}}

@inproceedings{Salvo20,
    author={Salvo, Rosa Di and Fazio, Maria and Celesti, Antonio and Santoro, Domenico and Villari, Massimo},
    booktitle={2020 IEEE Globecom Workshops (GC Wkshps}, 
    title={Mathematical Model and AI Oriented Analysis for Self-Regulated Learning in Remote Health Treatments}, 
    year={2020},
    publisher    = {IEEE},
    address      = {New York, NY, USA},
    pages={1-6},
    doi={10.1109/GCWkshps50303.2020.9367487}
}

@inproceedings{Afzaal21,
    author={Afzaal, Muhammad and Nouri, Jalal and Zia, Aayesha and Papapetrou, Panagiotis and Fors, Uno and Wu, Yongchao and Li, Xiu and Weegar, Rebecka},
    title={Automatic and Intelligent Recommendations to Support Students’ Self-Regulation},
    booktitle={2021 International Conference on Advanced Learning Technologies (ICALT)},
    year={2021},
    pages={336-338},
    publisher    = {IEEE},
    address      = {New York, NY, USA},
    doi={10.1109/ICALT52272.2021.00107}
}

@inproceedings{Yang24,
    author={Yang, Qin and Lu, Guoqing and He, Xiangchun and Zhang, Chenwen},
    booktitle={2024 International Symposium on Educational Technology (ISET)}, 
    title={How Pre-service Teachers’ Basic Need Satisfaction Affect their AI Literacy in AI Introductory Courses? the Roles of Emotional Engagement and Self-regulated Learning}, 
    year={2024},
    publisher    = {IEEE},
    address      = {New York, NY, USA},
    pages={103-108},
    doi={10.1109/ISET61814.2024.00029}
}

\end{document}